\documentclass[aps,pre,twocolumn,showpacs,superscriptaddress,groupedaddress]{revtex4}
\usepackage{graphicx,subfigure}
\usepackage{hyperref}
\usepackage[svgnames]{xcolor}
\usepackage{amsmath}
\usepackage{amsfonts}
\usepackage{bm}
\usepackage{amssymb}
\usepackage{varioref}
\usepackage{float}
\usepackage{dcolumn}   
\usepackage{subfigure}
\usepackage{array}
\usepackage{tikz}
\usepackage{pgfplots}
\usepackage{mathtools}
\usetikzlibrary{positioning,fadings,through}
\definecolor{left} {HTML}{001528}

\begin{document}

\title[Spiral folding of a flexible chain of chiral active particles]{Spiral folding of a flexible chain of chiral active particles}

\author{Shalabh K. Anand}

\affiliation{Department of Bioengineering, Imperial College London, South Kensington Campus, London SW7 2AZ, United Kingdom}
\affiliation{Department of Mathematics, Imperial College London, South Kensington Campus, London SW7 2AZ, United Kingdom}
\affiliation{Aix Marseille Univ, Université de Toulon, CNRS, CPT (UMR 7332), Turing Centre for Living systems, Marseille, France}

\email{shalabh.kumar.anand@gmail.com}
\vspace{10pt}

\begin{abstract}
We investigate a flexible polymer chain made up of chiral active Brownian particles in two dimensions using computer simulations. In the presence of chiral active Brownian forces, the radius of gyration of the chain reduces significantly. We further identify the formation of spirals using the tangent-tangent correlation to characterize the internal structure of the chain. The polymer chain forms a pair of spirals with opposite spiral turns on both ends of the polymer. We compute the number of turns of both spirals, and find that the total number of turns increases with angular frequency as well as P{\'e}clet number. However, the spirals become weak and the number of turns decreases at a very high P{\'e}clet number. We draw a phase diagram using the turn number. The end-to-end correlation displays oscillatory behavior, which signifies the rotational dynamics of the chain. We quantify the rotation frequency from the end-to-end vector, which follows a power law behavior with exponent $3/2$. We also provide a scaling relation between the radius of gyration and the chain length, and the exponent decreases significantly in the presence of chiral active forces.
\end{abstract}

\maketitle

\section{Introduction}
Active matter systems have gathered significant research interest over the past few decades due to their potential to provide quantitative insights into various biological processes~\cite{marchetti2013hydrodynamics,ramaswamy2010mechanics,winkler2017active}. These systems are typically driven out of equilibrium by the input of energy, which is then converted into mechanical work. As a result, they exhibit behavior that is distinctively different from their passive counterparts, such as motility-induced phase separation (MIPS), pattern formation~\cite{vicsek1995novel, toner1995long}, and other emerging phenomena~\cite{elgeti2015physics, bechinger2016active,anand2024active,juelicher2007active}. Active matter systems encompass a wide range of entities, from naturally occurring macroscopic and microscopic objects to particles synthesized in artificial environments~\cite{yan2016reconfiguring,walther2013janus,arlt2018painting,deblais2023worm}.

Among the diverse active matter systems, this manuscript focuses on active polymers~\cite{shee2021semiflexible,das2021coil,anand2019beating,clopes2021flagellar,bianco2018globulelike,fily2020buckling}. These polymers are abundant in nature, with examples such as microtubules and actin filaments found within cells~\cite{harada1987sliding,heeremans2022chromatographic,manna2019emergent,brennen1977fluid}. Microtubules and actin filaments, which are components of the cytoskeleton in living cells, exhibit directed motion facilitated by motor proteins~\cite{howard2002mechanics,ndlec1997self}. Over the past decade, these filaments have been extensively studied, both experimentally and theoretically~\cite{nishiguchi2018flagellar,deblais2020phase,anand2019behavior,biswas2017linking}, due to their intriguing structural~\cite{anand2018structure,anand2020conformation,vatin2024conformation,natali2020local}, dynamical~\cite{vandebroek2015dynamics,ghosh2014dynamics,prathyusha2018dynamically,laskar2017filament,chelakkot2014flagellar}, and collective properties~\cite{ravichandran2017enhanced,abkenar2013collective,foglino2019non}.

From a theoretical modeling perspective, active polymers are generally categorized into two main types. The first considers the monomers of the polymer as active Brownian particles~\cite{eisenstecken2016conformational,eisenstecken2017internal,samanta2016chain,paul2022activity,das2019deviations}, while the second focuses on tangential activity, classifying the polymer as polar due to its unidirectional activity~\cite{anand2018structure,philipps2022tangentially,tejedor2024progressive,fazelzadeh2023effects,malgaretti2024coil,lamura2024excluded}. Both types of activity result in intriguing, yet distinct behaviors. Active Brownian forcing induces non-monotonic structural changes, whereas tangential activity leads to significant shrinkage of the polymer chain. In two-dimensional systems, tangentially propelled polymer chains adopt a spiral conformation~\cite{isele2015self,gupta2019morphological,D4SM00511B}, a feature absent in polymers governed by active Brownian forces~\cite{kaiser2015does,khalilian2024structural,chaki2019enhanced,goswami2022reconfiguration}.

In recent years, there has been growing research interest in the study of chiral active systems, revealing a range of novel behaviors that are not observed in achiral systems~\cite{oliver2018synchronization,zhang2020reconfigurable,kreienkamp2022clustering,liu2019configuration,li2023chirality}. This increased focus is largely driven by the fact that many naturally occurring active agents are also chiral~\cite{liebchen2022chiral,lauga2006swimming}. Chirality can arise from various factors, such as structural asymmetry, self-organization, or interactions among agents~\cite{kraft2013brownian,aubret2018targeted,vutukuri2017rational,wykes2016dynamic,narinder2018memory,kummel2013circular}. The presence of chirality introduces an additional angular velocity, which has been found to play a crucial and, in some cases, decisive role in determining the behavior of many systems~\cite{caprini2019active,arora2021emergent,caprini2023chiral}. In recent years, a polymer in the bath of chiral active particles~\cite{liu2019configuration,zhou2023dynamics}, and also a chiral active ring polymer~\cite{anand2024computer} has been studied. However, a chiral active linear polymer has not been studied to the best of our knowledge. In this manuscript, our aim is to provide a detailed analysis of a chiral active Brownian flexible polymer chain.

 The manuscript presents a comprehensive study of an active Brownian polymer under the influence of chirality in two dimensions. The active forces are imposed in random directions along with the rotational diffusion and an intrinsic angular frequency, which provides chirality to the monomers. We observe a formation of a pair of spirals at both ends of the polymer. The presence of spirals in a polymer chain is observed only in the presence of tangential activity, but we obtain a distinctive spiral formation in a pair in an active Brownian polymer chain because of the chiral forces. We quantify the total number of turns and also the difference between the number of turns on both ends. We obtain an increase in number of turns as a function of strength of the active force followed by a decay at high active force. On the basis of the spiral turns, we draw a phase diagram in the parameter regime where the polymer forms spirals. Along with the spirals in pairs, the polymer undergoes rotational motion. We identify this using time correlation of end-to-end vector, and we also compute the rotational frequency that follows a power-law behavior with active forces with exponent $3/2$. The diffusion of the polymer follows the same trend as that of an active Brownian chain, but it gets suppressed with angular velocity. We also elucidate the structural properties of the chain, which suggests the shrinkage of the chain with chirality at a given active force strength. 

The manuscript is organized as follows. First, we provide the modeling of the chiral active Brownian polymer in Section II. Then, we identify the conformations attained by the polymer in the presence of chiral active forces in Section III. We also discuss its structural and dynamical properties in detail. In the end, we summarize our study in Section IV.

\section{Model}
\begin{figure}
    \centering
    \includegraphics[width=\columnwidth]{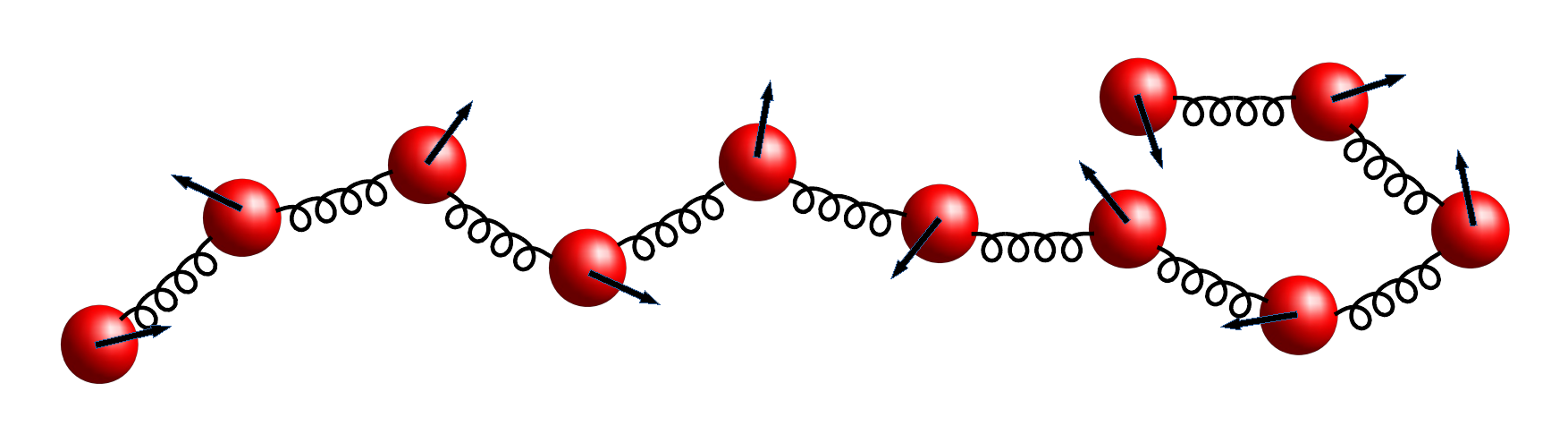}
    \caption{ A schematic to demonstrate the model of an active Brownian chain. Arrows in the figure denotes the directions of the active forces of the corresponding monomers.}
    \label{Fig:1}
\end{figure}
We model a flexible polymer as a linear chain of $N$ circular monomer beads connected through harmonic springs between successive monomers~\cite{lamura2022self,kaiser2015does,anand2021migration}. The spring potential energy $U_{h}$ is written as,
\begin{equation}
    {U_h} = \frac{k_h}{2} \sum_{i=1}^{N-1} (|\bm{r}_{i+1} - \bm{r}_i| - l_0)^2.
    \label{Eq:spring}
\end{equation}
Here $k_h$ is the spring constant, $l_{0}$ is the rest length of the spring, $\bm{r}_i$ is the position of $i^{th}$ monomer. Along with the spring potential, monomers interact among themselves via the truncated Lennard-Jones potential to avoid overlapping. Thus, the interaction energy for the inter-monomer distance between monomers $i$ and $j$, $R_{ij} < 2^{1/6} \sigma$, is,
\begin{equation}
U_{LJ} = 4 \epsilon \left[\left(\frac{\sigma}{R_{ij}}\right)^{12}- \left( \frac{\sigma}{R_{ij}}\right)^{6} \right] + \epsilon.
\end{equation}
For $R_{ij} \geq 2^{1/6}\sigma$, $U_{LJ}=0$, where ${\bm R}_{ij}={\bm r}_j-{\bm r}_i$,  $\epsilon$ is interaction energy and $\sigma$ is the diameter of the monomer.

The dynamics of the polymer chain is governed by the overdamped Langevin equation,
\begin{equation}
    \gamma \bm{\dot{r}_{i}}(t) = -\nabla_i{U}(t)  + \bm{\eta}_{i}^{transl}(t) + F_{a}\hat{\textbf{u}}_{i}(t).
    \label{Eq:eq_motion}
\end{equation}	
Here $U = U_{h}+U_{LJ}$ is the total interaction potential, $\gamma$ is the drag coefficient, $\bm{\eta}_{i}^{transl}(t)$ is the thermal noise on $i^{th}$ monomer and $F_{a}$ is the strength of the active force in the direction denoted by unit vector $\hat{\textbf{u}}_{i} = (cos\theta_{i}, sin\theta_{i})$. The angle $\theta_{i}$ is measured from x-axis for each monomer $i$, and evolves according to
\begin{equation}
\dot{\theta_{i}}(t) =  \eta_{i}^{rot}(t) + \omega.
\label{eq:rotlangevin}
\end{equation}
Here $\eta_{i}^{rot}$ is the random torque and $\omega$ is the angular frequency. 

The thermal noise $\eta_{i}^{transl}$ and the random torque $\eta_{i}^{rot}$ both have zero mean and their second moment follow the following relations,
\begin{eqnarray}
    \langle \bm{\eta}_{i}^{transl}(t) \cdot \bm{\eta}_{j}^{transl}(t') \rangle &=& 4k_B T \gamma \delta_{ij} \delta(t-t'), \\
    \big \langle \eta_{i}^{rot}(t).\eta_{j}^{rot}(t')\big \rangle &=& 2D_{rot}\delta(t-t')\delta_{ij}.
\end{eqnarray}

We present the strength of the active force as a dimensionless quantity P{\'e}clet number $Pe$ defined as a ratio of the active force with the thermal force, $Pe = (F_{a} l_0)/(k_{B}T)$. A schematics of an active Brownian polymer is displayed in Fig.~\ref{Fig:1}, where the arrows show the direction of active forces on each monomer. To characterize chirality, we define a dimensionless parameter $\Omega$, which is the ratio of deterministic and stochastic angular frequencies as $\Omega=\omega/D_{rot}$.

We non-dimensionalize the equation of motion using basic units such as bond length ${l_{0}}$ , diffusion coefficient of a monomer ${D_{transl}}$, and thermal energy ${k_{B} T}$. We present all the physical parameters in this manuscript in terms of these basic units. The number of monomers in a ring is fixed to be $N=200$ unless otherwise specified. The simulation parameters are chosen as ${k_{h}} = 1000\frac{k_{B}T}{l_{0}^{2}}, \sigma = {l_{0}}, \frac{\epsilon}{k_{B}T} = {1}$, and time is in units of $\tau = \frac{l_{0}^{2}}{D_{transl}} $. The rotational diffusion coefficient is taken as $D_{R}=3D_{transl}/\sigma^{2}$. We have performed all the simulations in a two-dimensional periodic box. We use the Euler integration method for the integration of the equation of motion with a time-step range between${10^{{-}4}\tau}-10^{-5}\tau$ to ensure the stability of the simulation. To ensure good statistics, each result is averaged over 10 different simulations with independent initial configurations. Once the system is equilibrated we compute the quantities presented in the manuscript.

\section{Results}
An active Brownian polymer shrinks and then swells with the strength of the active force~\cite{kaiser2015does,anand2020conformation}. A similar behavior is observed for a passive polymer chain in a bath of active particles~\cite{kaiser2014unusual,harder2014activity,anderson2022polymer}. However, a passive polymer chain in the bath of chiral active particles exhibits distinctive configurations compared to those of the achiral active bath. In this section, we present the structural and dynamic behavior of a flexible polymer chain in the presence of chiral active forces on its monomers. We systematically vary the strength of the active force and the magnitude of the angular frequency and analyze the conformations and dynamics of the chain.

\subsection{Structural properties}
\begin{figure*}
    \centering
    \includegraphics[width=1.02\textwidth]{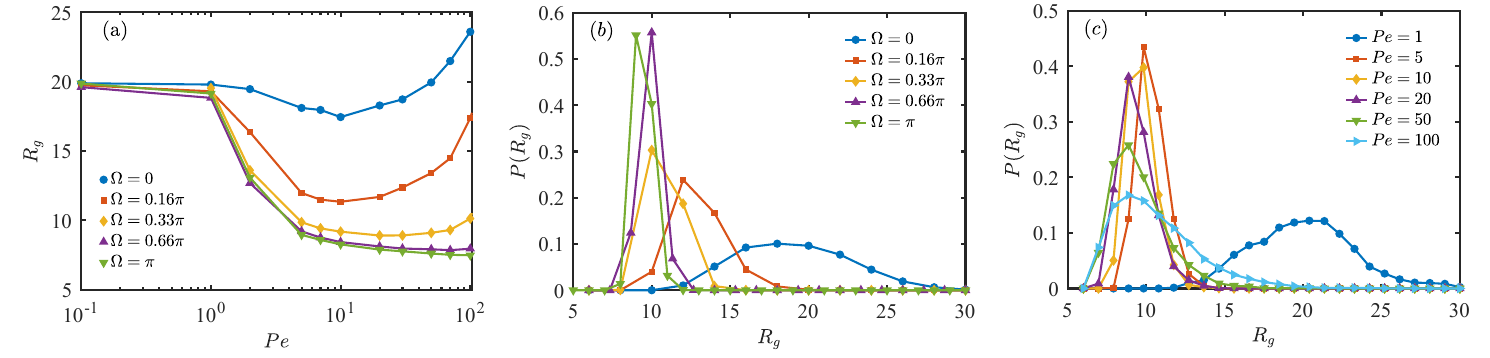}
    \caption{a) Average radius of gyration $R_g$ as a function of $Pe$ for various values of angular frequencies $\Omega$. b) Distribution of radius of gyration of the polymer for $Pe=5$ for different $\Omega$.  c) Distribution of radius of gyration of the polymer for $\Omega=0.33\pi$ at different $Pe$ strengths.}
    \label{Fig:structure}
\end{figure*}
\subsubsection{Radius of gyration--}
The structure of the polymer chain is discussed using the radius of gyration $R_{g}$  defined as $R_{g} = \sqrt{\langle\frac{1}{N}\sum_{i=1}^{N}(\bm{r}_{i}-\bm{r}_{cm})^{2}\rangle}$. Figure~\ref{Fig:structure}-a shows the average value of radius of gyration in the parameter space of $Pe$ and $\Omega$. For the achiral case, i.e. $\Omega=0$, the chain becomes an active Brownian polymer, and its $R_{g}$ varies non-monotonically with shrinkage in the intermediate $Pe$ regime followed by swelling at high $Pe$ strengths. Moreover, for $\Omega > 0$, the behavior remains unchanged in the low $Pe$ regime. However, $R_{g}$ decreases significantly in the intermediate $Pe$ regime. With  further increasing the strengths of $Pe$, the $R_{g}$ increases akin to an active Brownian polymer. Thus, the trend of variation of $R_{g}$ remains same for a chiral active polymer, i.e. for nonzero values of $\Omega$. The nonmonotonic behavior of $R_{g}$ remains consistent for chiral active forces, but with a significant shrinkage in $R_{g}$ with $\Omega$ at a given $Pe$ strength. Thus, the chiral active Brownian polymer chain shrinks a lot compared to its achiral counterpart. This trend persists even in the absence of excluded-volume interactions (see supplement~\cite{Note1} Fig.~SI-1). In addition, the shrinkage intensifies with $\Omega$ at a fixed $Pe$ initially before reaching to almost saturation at high $\Omega$.

The conformational response of an active Brownian flexible polymer to the variation in $Pe$ is non-monotonic, which contrasts with an active Rouse chain that shows a steady increase in $R_{g}$ with $Pe$~\cite{anand2020conformation,kaiser2015does}. The shrinkage of an active Brownian flexible polymer in the intermediate regime of $Pe$ is attributed to the excluded-volume effect, which disappears and shows a monotonic increase in the absence of excluded-volume interactions~\cite{anand2020conformation} (see supplement~\cite{Note1} Fig.~SI-1). When chiral active forces are introduced, the direction of the active force rotates with the given angular velocity. The neighboring monomers are connected via springs and it constrains them to stay together. Eventually, the monomers tend to fold the polymer chain locally, and the chain shrinks as a result. The high value of angular frequency causes a faster rotation, which leads to a more local folding, and consequently the chain forms a more compact structure. Thus, shrinkage of the polymer chain occurs primarily as a result of the chiral active forces. At a very high $Pe$, the monomers move very fast. Thus, the formation of compact structures becomes a difficult event in the presence of chirality, leading to an increase in $R_g$ in the large $Pe$ regime.

Furthermore, the role of chirality in the structural changes in the polymer is explored by analyzing the distribution of $R_{g}$ at a fixed active force $Pe=5$ for various values of $\Omega$ (see Fig.~\ref{Fig:structure}-b). The distribution is broad with a peak for $\Omega=0$, which becomes narrow with the peak shifting towards a smaller value of $R_{g}$ for chiral cases, i.e. $\Omega>0$. Further increasing the $\Omega$ moves the peak towards a much smaller value of $R_{g}$, reinforcing the chain shrinkage with $\Omega$ at a fixed $Pe$. In addition, the peak becomes narrow with $\Omega$, suggesting that the conformations attained in the parameter regime are very stable. We also examine how chirality influences polymer conformations across different $Pe$ values by analyzing the distribution of $R_g$ at $\Omega=0.33\pi$ (see Fig.~\ref{Fig:structure}-c). The chain shrinks in the intermediate regime as the peak is shifted to a smaller $R_g$ and also narrows. However, the peak becomes broad at high $Pe$ values, suggesting that the chain attains shrinked as well as open structures in this regime (see snapshots in supplement~\cite{Note1} Fig.~SI-4 a-e). This explains the observed swelling of $R_g$ at high $Pe$.

\begin{figure}
    \centering
    \includegraphics[width=\columnwidth]{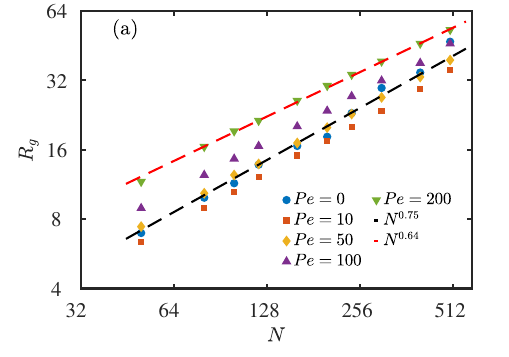}
    \includegraphics[width=\columnwidth]{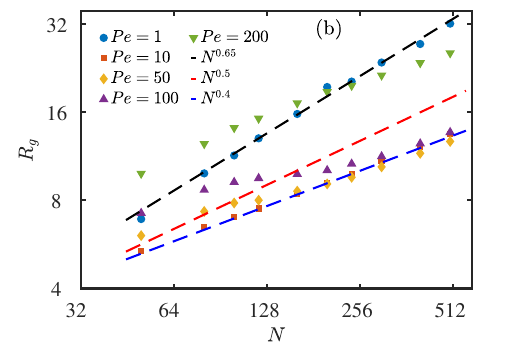}
    \caption{Scaling of radius of gyration $R_{g}$ with chain length $N$ for a) $Pe=0, 10, 50, 100$ and $200$ at $\Omega=0$ and b) $Pe=1, 10, 50, 100$ and $200$ at $\Omega=\pi$. Dotted lines are power-law fitting with the given exponents.}
    \label{Fig:scaling}
\end{figure}
We now examine the impact of chain shrinkage in the presence of $\Omega$ on the scaling relation between $R_{g}$ and the number of monomers $N$. First, we present the scaling between $R_{g}$ and $N$ for an active Brownian chain, i.e. $\Omega=0$ (see Fig.~\ref{Fig:scaling}-a). As expected, the scaling relation for a self-avoiding chain at $Pe=0$ follows $R_{g} \sim N^{\nu}$ with $\nu = 3/4$~\cite{kaiser2015does}. The exponent of the power-law relation does not change much at a small $Pe$. However, the exponent decreases significantly ($\nu \approx 0.64$) at a high active force ($Pe = 200$).

We now discuss the effect of chirality on the scaling exponent mentioned above. Figure~\ref{Fig:scaling}-b displays the power-law behavior of $R_{g}$ with the number of monomers $N$ at $\Omega=0.33\pi$ for various strengths of active forces $Pe$. We observe that even at a small active force, say at $Pe=1$, the exponent is $\nu \approx 0.65$ which is significantly smaller than that for $\Omega=0$. We further increase $Pe$, and the exponent $\nu$ also decreases as a consequence. In the intermediate regime of $Pe$, where $R_{g}$ hits its minima, the exponent is $\nu \approx 0.4$, which is well below the exponent exhibited by an ideal chain ($1/2$). These are the conformations with significantly reduced structural quantities $R_g$ and $R_e$. However, at a very high $Pe$, the exponent varies in two steps since $\nu$ is different in the short- and long-length regimes. In the short-length regime, the exponent is slightly higher, and then it decreases to a much smaller value in a higher polymer length regime. The different behaviors in short and long polymer chains can be understood as follows: The monomers rotate because of the chiral active forces and try to fold locally. However, it is easy to escape the monomers from the folded structures when the chain length is not high. So, the polymer chain keeps changing its structure from closed to open and vice versa. Thus, the average value of $R_{g}$ is high. However, a long chain forms a compact structure, resulting in a smaller scaling exponent in the long-length regime.

\begin{figure}
    \centering
    \includegraphics[width=\columnwidth]{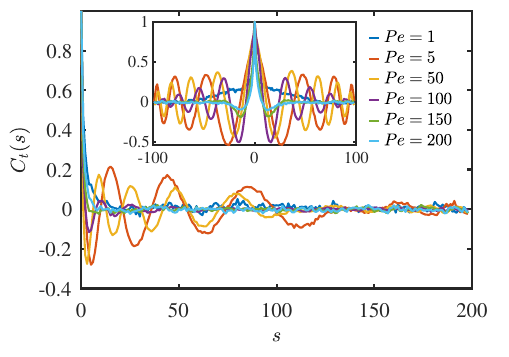}
    \caption{Bond correlation starting from the front to the rear end at $\Omega=\pi$. Inset displays the correlation from middle of the chain to both ends for the same parameters.}
    \label{Fig:6}
\end{figure}
 We further investigate the spatial correlation along the contour of the polymer using the bond correlation defined as $C_{b}(s) = \langle \bm{t}(s).\bm{t}(0) \rangle$. Here, $\bm{t}(s)$ is the unit bond vector, i.e. $(\bm{r}_{i+1}-\bm{r}_{i})/|\bm{r}_{i+1}-\bm{r}_{i}|$. Figure~\ref{Fig:6} displays the correlation function $C_{b}(s)$ from one end to the other at a fixed angular frequency $\Omega=\pi$. At a small active force $Pe=1$, the correlation quickly decays to zero. However, as we increase $Pe$ further, the correlation exhibits oscillations along the contour $s$. The oscillatory bond correlation suggests that there are local foldings that are correlated along the chain; thus it can be a sign of spiral formation. However, the oscillation stops almost mid-way along the polymer contour and decays to zero. The oscillations in the correlation function also stop at high active forces. To understand the oscillatory tangent correlation up to mid-way along the chain, we compute the correlation function from the middle of the chain towards both ends (see inset of Fig.~\ref{Fig:6}). Here, the negative and positive arc distances are along the different ends of the chain. We observe an oscillatory correlation function on both sides of the chain. Thus, correlated folding occurs at both ends of the chain. However, they become decorrelated almost midway along the chain. 

\subsubsection{Spiral formation--}
\begin{figure}
    \centering
    \includegraphics[width=0.9\columnwidth]{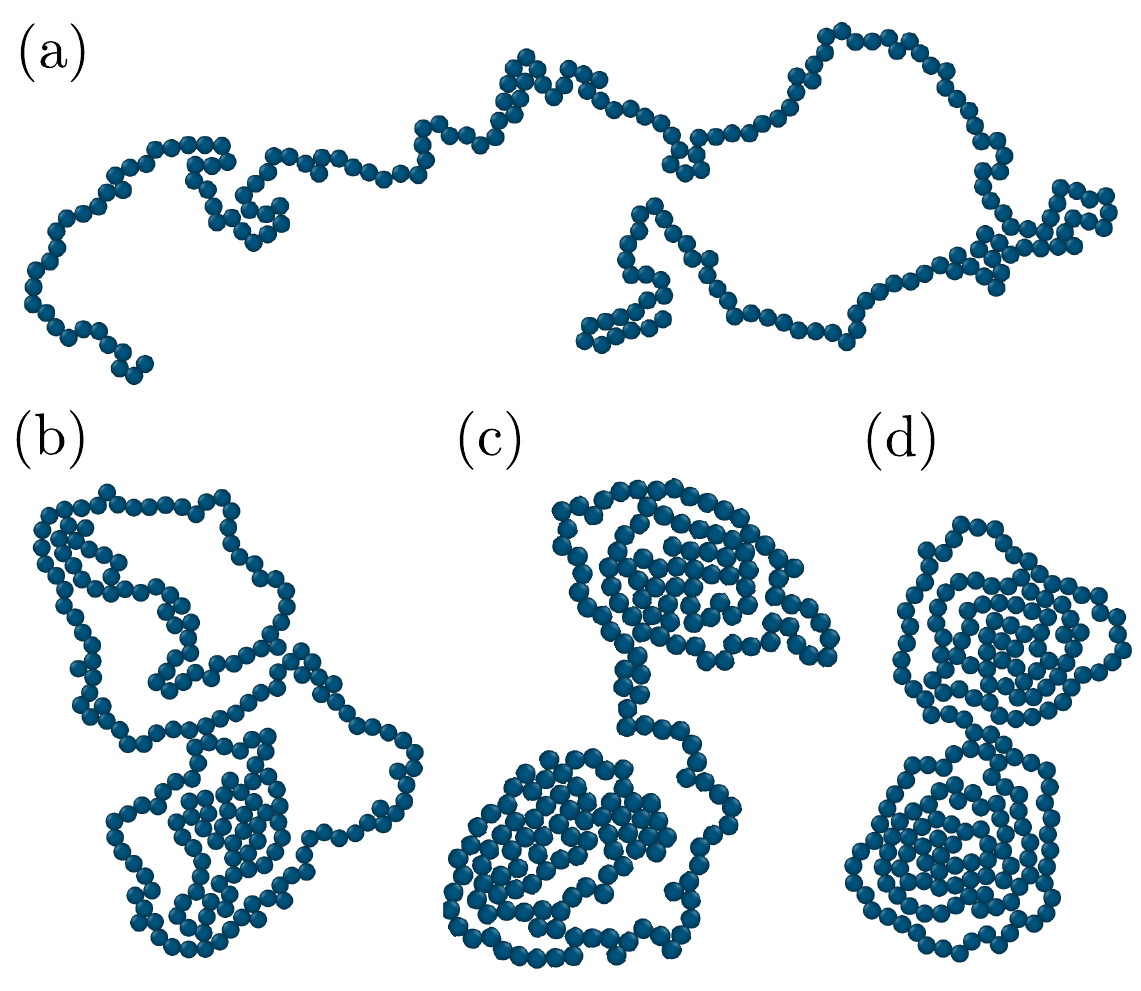}
    \caption{A few snapshots of the configurations attained by the polymer chain at $Pe=10$ for a) $\Omega=0$, b) $0.16\pi$, c) $0.33\pi$ and d) $\pi$.}
    \label{Fig:snaps}
\end{figure}
 As observed in the bond correlation function, a chiral active Brownian polymer forms a correlated folding at both ends of the chain. For visual clarity, we show a few snapshots of the chain conformations in Fig.~\ref{Fig:snaps} at $Pe=10$ for three different values of $\Omega$. The chain attains a random configuration at $\Omega=0$, however, it starts showing spiral turns at both ends when the angular frequency is tuned on, as for example at $\Omega=0.17\pi$. Moreover, the spiral turns at both ends seem weak. The number of turns start increasing with increasing $\Omega$ (see Fig.~\ref{Fig:snaps}-c and d at $\Omega=0.33\pi$ and $\pi$ respectively). We now quantify the spiral shapes using the parameter defined as turn number~\cite{shee2021semiflexible,isele2015self,sabeur2008kinetically,mitra2023emergent} as,
\begin{equation}
    \psi_{n} = \frac{1}{2\pi}\sum_{i=1}^{N-1}(\theta_{i+1}-\theta_{i}).
    \label{Eq:turn_num}
\end{equation}
 Here, $(\theta_{i+1}-\theta_{i})$ is the angle between two consecutive bonds. The quantity $\psi_{n}$ gives us the number of turns taken by the polymer chain. A negative value of $\psi_n$ corresponds to clockwise looping and a positive value of $\psi_n$ corresponds to counterclockwise looping of the chain, and it becomes zero for a straight configuration. Since the two spirals are separate and at the two different ends, we extract the number of turns into two parts: we start our computation from one end ($i=1$) to the middle of the chain and we name it $\psi_{1}$, then we compute the number of turns at the other end ($\psi_{2}$) starting from middle to $i=N$. This provides us a good measure, as the oscillations in the bond correlation function decay to zero until it reaches almost the mid-way ($s=100$) along the contour (see Fig.~\ref{Fig:6}).

\begin{figure}
    \centering
    \includegraphics[width=\columnwidth]{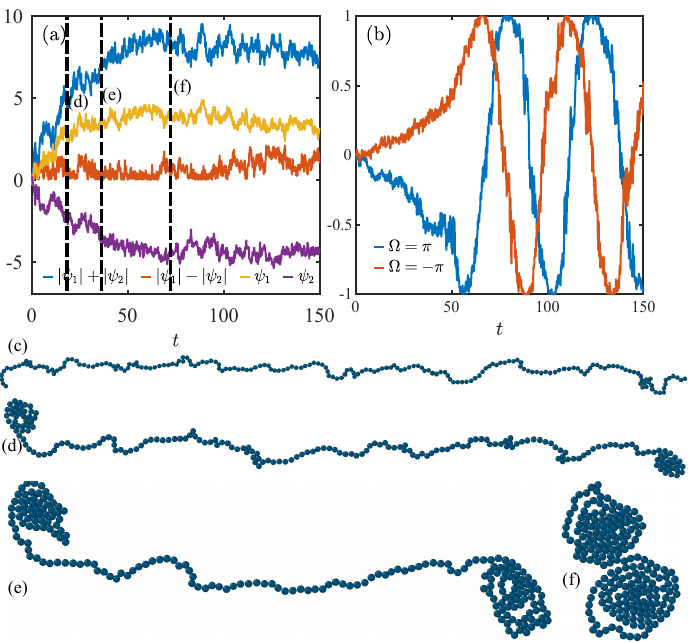}
    \caption{a) Time-series of number of spirals $\psi_L$, $\psi_R$, total number of spirals $|\psi_L|+|\psi_R|$ and the difference $|\psi_L+\psi_R|$. b) Time-series of x-component of end-to-end vector for clockwise and counter-clockwise chiral active forces. Snapshots at t=0 and time-points marked in figure a) are shown in c), d), e) and f). The strength of active force used in the simulations for this figure is $Pe=50$ at $\Omega=\pi$. }
    \label{Fig:time-series}
\end{figure}

We plot the time series of both $\psi_1$ and $\psi_2$ as well as the sum ($|\psi_1| + |\psi_2|$) and difference ($|\psi_1| - |\psi_2|$) of both turns in Fig.~\ref{Fig:time-series}-a. We start with an extended configuration, which is displayed in Fig.~\ref{Fig:time-series}-c. The chain forms a spiral with time, thus the number of turns increases on both sides, and so does the sum of turns on both sides. The turns at both ends are opposite in sign, thus a pair with opposite handedness. We mark the time points and display the corresponding configurations in Fig.~\ref{Fig:time-series}-d, e and f. After an initial increase, the turn number reaches a steady-state value. We will discuss the average of the sum and the difference of both turns in detail in the following paragraphs. We also plot time series of x-component of the end-to-end vector to find the dynamics. We observe that the chain rotates as the formation of the spirals occurs. We changed the chirality to a negative value and observed the same but with opposite direction of rotation. We will discuss the rotation of the chain later in this manuscript.

\begin{figure}
    \centering
    \includegraphics[width=\columnwidth]{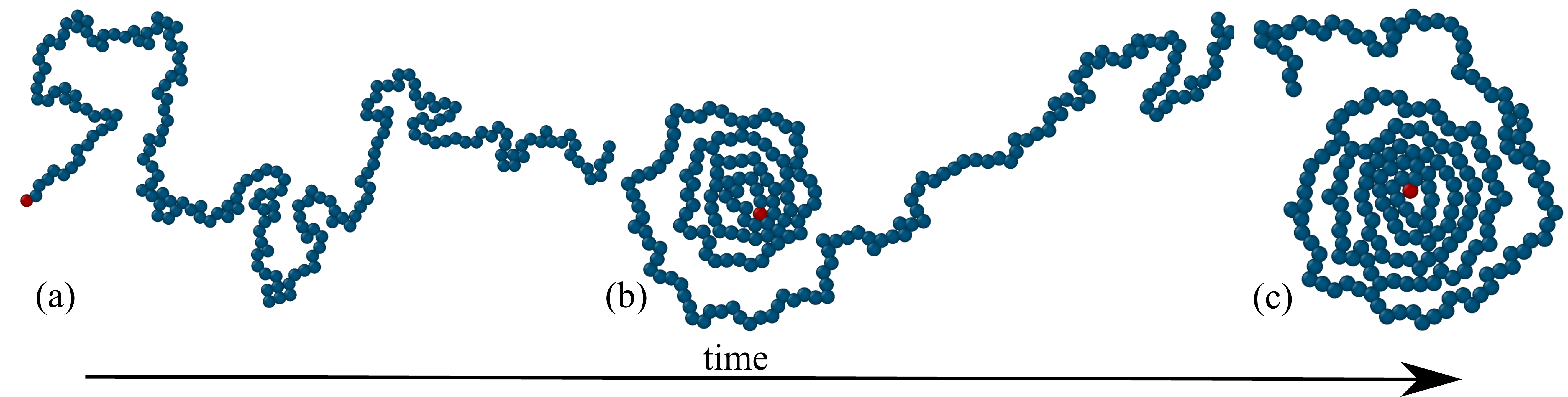}
    \includegraphics[width=\columnwidth]{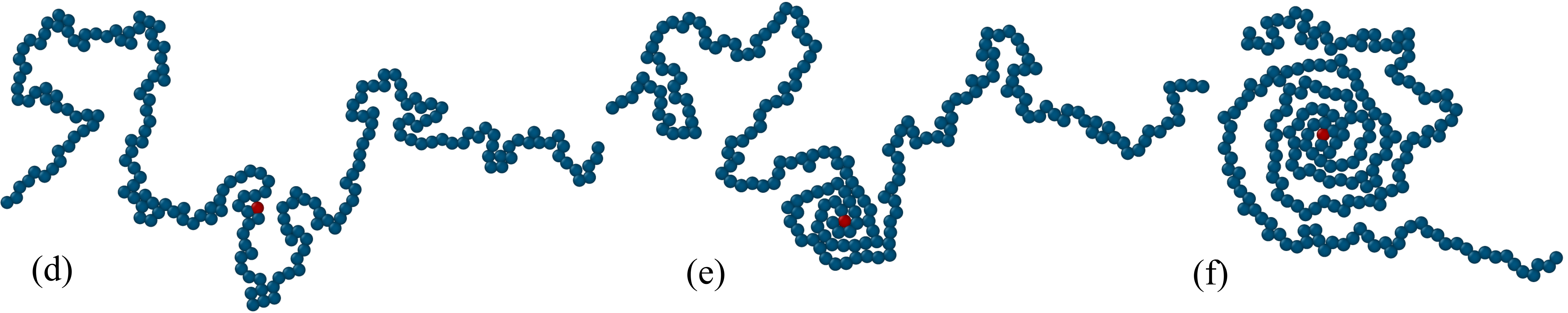}
    \caption{Snapshots showing time evolution of a polymer configuration with only one active monomer (passive monomers in blue and active in red). a) and d) are initial configuration with end active monomer and middle active monomer respectively. b), c) and e), f) are time evolution of configurations a) and d) respectively. The strength of active force used in the simulations for this figure is $Pe=50$ at $\Omega=\pi$.} 
    \label{Fig:snaps-head}
\end{figure}
To understand the mechanism behind this spiral formation, we consider a passive chain with one chiral active monomer. We consider two possibilities of placing the active monomer on the backbone of the chain: i) the end monomer is active, and ii) the middle monomer is active. Figure~\ref{Fig:snaps-head} shows the snapshots of these two cases in the long time limit for $Pe=50$ and $\Omega=\pi$. In both cases, we see that the chain forms a spiral around the active monomer. Thus, a single chiral active monomer can lead to spiral formation in a polymer. However, in this manuscript, we have a polymer chain made up of chiral active monomers. Here, all of the monomers try to fold the chain around themselves. A similar study on a ring polymer~\cite{anand2024active} does not form a spiral as it does for an open chain. On the other hand, an anchored chain of chiral active particles at one end forms one spiral (see supplement ~\cite{Note1} Fig.SI-4f and also in a recent preprint~\cite{caprini2024spontaneous}). Thus, the formation of two spirals at both ends can also potentially be a result of an open chain together with the chiral active particles.

\begin{figure}
    \centering
    \includegraphics[width=\columnwidth]{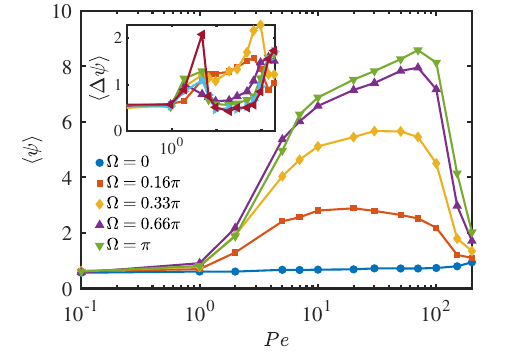}
    \caption{ Average number of total turns as a function of $Pe$ for various $\Omega$. Inset displays the difference in the number of spiral turns at both ends of the chain. }
    \label{Fig:spiral_num}
\end{figure}

We plot the average value of the sum of both turns as $\langle\psi\rangle = \langle|\psi_{1}| + |\psi_{2}|\rangle$ in Fig.~\ref{Fig:spiral_num}. As expected, we observe no spiral formation ($\langle\psi\rangle=0$) for achiral active Brownian polymer, that is, for $\Omega=0$. When the chirality is tuned on ($\Omega > 0$), a positive value of $\langle\psi\rangle$ develops with $Pe$ and the total number of turns increases with $Pe$. However, it decreases after an initial increase after a certain activity strength $Pe$. This trend remains consistent for higher $\Omega$ as well, and the total number of turns also increases with $\Omega$. The decrease in the total number of spirals at high $Pe$ is due to the opening of the folds, which we also see in the tangent correlation $C_{b}(s)$ in Fig.~\ref{Fig:6}. This opening of the spirals is also reflected in the $R_{g}$ at high $Pe$, where it increases with $Pe$. We also plot the distribution of both turns in the supplementary text~\cite{Note1} (Fig.~SI-3), which assists this observation.

We now investigate the difference in clockwise and counterclockwise turns on both sides, i.e. $\Delta\psi = |\psi_{1}| - |\psi_{2}|$. The inset of Fig.~\ref{Fig:spiral_num}-a displays $\Delta\psi$ as a function of $Pe$ for various angular frequencies $\Omega$. We observe that the difference is less than 1 for high $\omega$ and also in the intermediate range of $Pe$ where a pair of stable spirals are formed. However, the difference in the number of turns has increased to a higher value for the $Pe$ strengths at which the onset of spiral formation occurs. In this regime, sometimes monomers in the middle become part of one spiral and later they change to the another one. This is also reflected in the bond correlation (Fig.~\ref{Fig:6}), where the oscillations occur beyond the halfway of the chain. 

\begin{figure}
    \centering
    \includegraphics[width=\columnwidth]{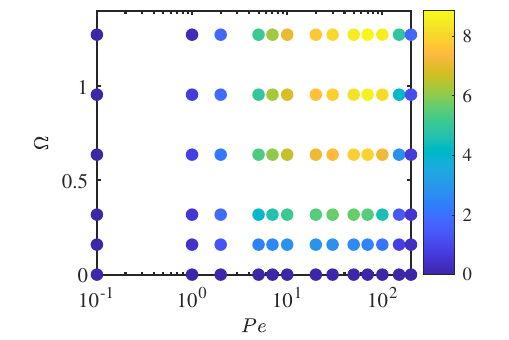}
    \caption{ Phase diagram of the spiral state in the space of $Pe$ and $\Omega$. Color bar displays the total number of turns $\langle\psi\rangle$.}
    \label{Fig:phase}
\end{figure}
So far, we have discussed the spiral formation and its opening in the parameter regime of $Pe$ and $\Omega$. Based on the total number of turns of the polymer, we draw a phase diagram in the parameter range of $Pe$ and $\Omega$ in Fig.~\ref{Fig:phase}. The color bar displays the total number of turns as displayed in Fig.~\ref{Fig:spiral_num}-a. It is clear that at small $Pe$ strengths, we obtain a polymer chain without spiral turns and also at very high $Pe$ strengths. However, we obtain a pair of spiral structures of the polymer in the intermediate range of $Pe$.

\subsection{Dynamics}
As discussed in the previous section, a chiral active Brownian polymer forms a pair of spiral folding at both ends of the chain. Now, we investigate whether the pair of spirals at both ends of the chain have implications on the dynamics as well.
\subsubsection{Relaxation --}
\begin{figure}
    \centering
    \includegraphics[width=\columnwidth]{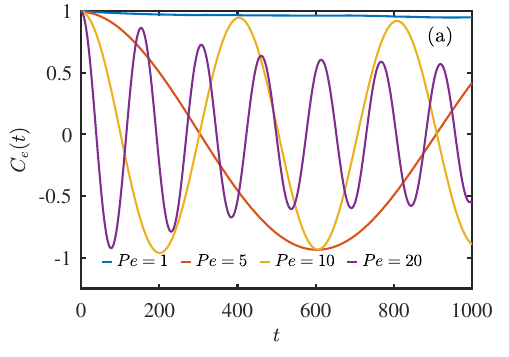}
    \includegraphics[width=\columnwidth]{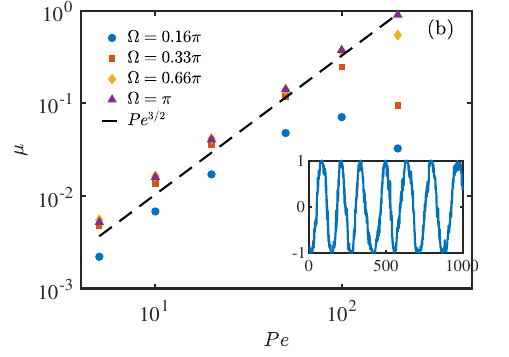}
    \caption{a) End-to-end correlation at $\Omega = \pi$. b) Frequency of rotation as a function of $Pe$. Inset displays the time-series data of x-component of end-to-end vector at $Pe=20$ for $\omega=2\pi$.}
    \label{Fig:8}
\end{figure}
We discuss the relaxation of the polymer chain using the time correlation function of the end-to-end vector, defined as $C_{e}(t) = \langle \bm{R}_{e}(t).\bm{R}_{e}(0) \rangle$. The end-to-end correlation of an equilibrium chain decays exponentially and provides us the relaxation time $\tau_{r}$ as $C_{e}(t) \sim exp(-t/\tau_{r})$. An active Brownian polymer chain decays faster than its counterpart passive chain, and the relaxation time becomes smaller~\cite{anand2020conformation,martin2020hydrodynamics,martin2019active} (see supplement~\cite{Note1} Fig.~SI-5). Here, we will discuss the correlation function for a chiral active chain. Figure~\ref{Fig:8}-a displays $C_{e}(t)$ as a function of time at $\Omega=\pi$ for various strengths of $Pe$. The correlation decays exponentially at a small $Pe$, which shows oscillations with further increase in $Pe$. The time period of oscillations decreases with further increase in $Pe$. However, the oscillatory behavior of the correlation function is suppressed at a very high $Pe$. The oscillations in end-to-end vector signs about the rotation of the chain with time. Usually, the oscillatory end-to-end vector correlation appears in the presence of external fields such as shear~\cite{liu2018individual,singh2020flow}, however, we observe the oscillations even in the absence of an external field. It is completely because of the intrinsic angular frequency. 

Moreover, we investigate the rotation of the polymer chain in the presence of chiral forces. We quantify the rotation using the time series data of the x-component of the end-to-end vector, which has a sinusoidal oscillations with time (see inset of Fig.~\ref{Fig:8}-b). We compute the frequency of the rotation by Fourier transform of this time series data. We extract the rotation frequency $\mu$ from the peak of the power-spectrum. Figure~\ref{Fig:8}-b displays the computed frequency of rotation as a function of $Pe$ for various magnitudes of $\Omega$.  The rotation frequency increases with $Pe$, and it follows a power law behavior with exponent $3/2$, i.e. $\mu \sim Pe^{3/2}$ similar to a chiral active ring polymer~\cite{anand2024computer}.

\subsubsection{Diffusion--}
\begin{figure}
    \centering
    \includegraphics[width=\columnwidth]{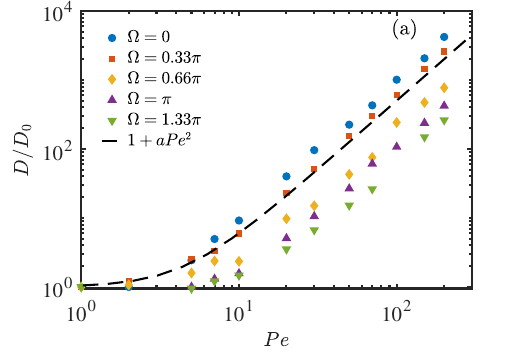}
    \includegraphics[width=\columnwidth]{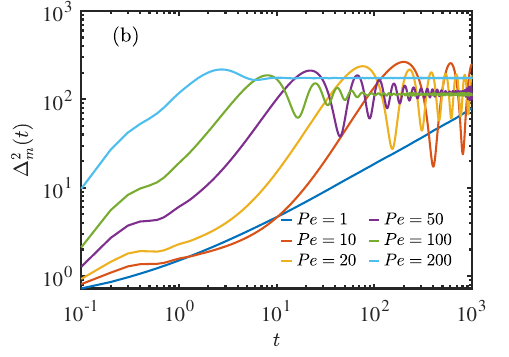}
    \caption{a) Diffusion coefficient as a function of $Pe$ for various $\Omega$. b) Mean-squared displacement of the monomers at $\Omega=\pi$ for various $Pe$.}
    \label{Fig:7}
\end{figure}
To analyse the polymer's dynamics, we calculate the mean-squared displacement (MSD) of its centre-of-mass (COM). The MSD of the COM, $<R_{cm}^{2}(t)> = \langle (\bm{r}_{cm}(t)-\bm{r}_{cm}(0))^{2}\rangle$, quantifies how far the COM moves over time. For an active Brownian polymer, the MSD exhibits a super-diffusive regime in short time-scale followed by a diffusive regime in the long time limit. We investigate how chirality affects the MSD. We characterize the dynamics of the polymer chain using diffusion coefficient calculated from the diffusive regime of the MSD from the relation $\langle R_{cm}^{2}(t) \rangle \sim 4Dt$. The inset of Fig.~\ref{Fig:7} illustrates the self-diffusion coefficient normalized by its value for a passive chain $D/D_{0}$. The diffusion of an active Brownian polymer varies as $D = D_{0}(1+a Pe^{2})$ with $Pe$~\cite{anand2020conformation}. As similar to a chiral active ring~\cite{anand2024computer}, we obtain the same relation for a linear chain as well. The fitting parameter $a$ changes with $\Omega$, and is listed in the table below. 

\begin{center}
    \begin{tabular}{ |c|c|c|c|c|c|c|c| }
        \hline
        $\Omega$ & 0 & $0.33\pi$ & $0.66\pi$ & $\pi$ & $1.33\pi$  \\
        \hline
        $a$  & 0.1014 & 0.0643 & 0.0198 & 0.01052 & 0.00652 \\
        \hline
    \end{tabular}
\end{center}

As we have discussed in the previous paragraphs, the chain rotates in the presence of chiral forces. Now, we move on to the internal dynamics using MSD of the monomers, that is, $\Delta_{m}^{2}(t)$. Figure~\ref{Fig:7}-b shows the MSD of the monomers averaged over all monomers from the centre-of-mass frame for various $Pe$ strengths at the chirality parameter $\Omega=\pi$. The MSD of the monomers of an active Brownian polymer initially exhibits super-diffusive behavior, which reaches to a plateau as time progresses. The plateau indicates that the monomers can no longer diffuse beyond the scale of polymer chain's length. Here, we will discuss the effects of chirality on the MSD. As expected, the short-time MSD exhibits super-diffusive behavior. Moreover, the MSD exhibits oscillations in the plateau regime, which is very similar to a chiral active ring~\cite{anand2024computer}. These oscillations suggest that the monomers undergo a rotational motion around their position in addition to translation. The onset of these oscillations occurs at a smaller time  for a larger  P\'eclet number. The oscillatory MSD aligns with the observation we have made earlier in this manuscript that the chain undergoes rotational motion.

\section{Conclusion} 
In summary, we have systematically studied how 
 the conformation and the dynamics of a linear polymer is influenced by chiral active Brownian forces. We observed the formation of a pair of spirals on both ends of the polymer chain with angular frequency. The spiral formation in active polymers has been found in the presence of tangential activity; however, the chirality causes the formation of a pair of spirals for the active Brownian forces. We quantified the number of spiral turns and found that the polymer entered into spiral phase in the intermediate regime of activity, which again broke at high active forces. Based on our observation, we have drawn a phase diagram in the parameter regime of $Pe$ and $\Omega$. 

We analyzed the structure of the chain using radius of gyration, and it decreased significantly compared to the achiral active Brownian polymer. This structural change is also mirrored in the internal structure of the chain, and the bond correlation exhibited oscillations. Using bond correlation, we showed the formation of correlated folding, which eventually led to shrinkage of the chain as a result of chiral active forces. We also presented a scaling exponent from the relation between the radius of gyration and the number of monomers. The exponent became significantly smaller in the presence of angular frequency.

We have also studied the impact of spiral formation on the dynamics of the polymer, and using the time correlation of end-to-end vector we showed that the polymer undergoes rotational motion and its rotational frequency follows a power-law behavior with exponent $3/2$. We further extracted the diffusion coefficient of the polymer and it varied with the same trend with $Pe$ as achiral active Brownian polymer but with a smaller diffusion coefficient. We have also shown that the spiral formation is led by chiral active monomers. However, a pair of spirals is also a result of an open linear chain. We have shown in our earlier work that a chiral active ring does not form spirals~\cite{anand2024computer}. Though, there is local folding of monomers in the ring which led to a shrinkage in the radius of gyration with chirality at a given activity strength. However, a ring doesnot form a spiral as a linear chain does, and the number of spirals increases with active force as well as with angular frequency.

Our study provided a detailed view of structural and dynamical changes in a polymer chain in the presence of chiral active forces. Chirality is very common in nature, thus, it becomes important to understand its impact on active biopolymers. In nature, we see many complex structures being formed and collapsed structures are also found. We showed a pair of spiral formations in our simulations; thus chirality can lead to many complex patterns. We have also explored the variation of the rotational diffusion constant to vary $\Omega$ in the supplementary text. We observed a similar pattern in the variation $R_g$ with $Pe$ and also the spiral formation (see Fig.~SI-2 in supplement~\cite{Note1}). The future direction related to our studies will be to explore the role of alignment interactions on an active Brownian polymer. 

\section{Acknowledgments}
We acknowledge the research computing facilities at Imperial College London.

\section{References}
\nocite{*}

\providecommand{\newblock}{}


\begin{thebibliography}{10}
\expandafter\ifx\csname url\endcsname\relax
  \def\url#1{{\tt #1}}\fi
\expandafter\ifx\csname urlprefix\endcsname\relax\def\urlprefix{URL }\fi
\providecommand{\eprint}[2][]{\url{#2}}


\bibitem{marchetti2013hydrodynamics}
Marchetti M~C, Joanny J~F, Ramaswamy S, Liverpool T~B, Prost J, Rao M and Simha R~A 2013 {\em Reviews of modern physics\/} {\bf 85} 1143

\bibitem{ramaswamy2010mechanics}
Ramaswamy S 2010 {\em Annu. Rev. Condens. Matter Phys.\/} {\bf 1} 323--345

\bibitem{winkler2017active}
Winkler R~G, Elgeti J and Gompper G 2017 {\em Journal of the Physical Society of Japan\/} {\bf 86} 101014

\bibitem{vicsek1995novel}
Vicsek T, Czir{\'o}k A, Ben-Jacob E, Cohen I and Shochet O 1995 {\em Physical review letters\/} {\bf 75} 1226

\bibitem{toner1995long}
Toner J and Tu Y 1995 {\em Physical review letters\/} {\bf 75} 4326

\bibitem{elgeti2015physics}
Elgeti J, Winkler R~G and Gompper G 2015 {\em Reports on progress in physics\/} {\bf 78} 056601

\bibitem{bechinger2016active}
Bechinger C, Di~Leonardo R, L{\"o}wen H, Reichhardt C, Volpe G and Volpe G 2016 {\em Reviews of Modern Physics\/} {\bf 88} 045006

\bibitem{anand2024active}
Anand S~K, Lee C~F and Bertrand T 2024 {\em Physical Review Research\/} {\bf 6} L022018

\bibitem{juelicher2007active}
Juelicher F, Kruse K, Prost J and Joanny J~F 2007 {\em Physics reports\/} {\bf 449} 3--28

\bibitem{yan2016reconfiguring}
Yan J, Han M, Zhang J, Xu C, Luijten E and Granick S 2016 {\em Nature materials\/} {\bf 15} 1095--1099

\bibitem{walther2013janus}
Walther A and Muller A~H 2013 {\em Chemical reviews\/} {\bf 113} 5194--5261

\bibitem{arlt2018painting}
Arlt J, Martinez V~A, Dawson A, Pilizota T and Poon W~C 2018 {\em Nature communications\/} {\bf 9} 768

\bibitem{deblais2023worm}
Deblais A, Prathyusha K, Sinaasappel R, Tuazon H, Tiwari I, Patil V~P and Bhamla M~S 2023 {\em Soft Matter\/} {\bf 19} 7057--7069

\bibitem{shee2021semiflexible}
Shee A, Gupta N, Chaudhuri A and Chaudhuri D 2021 {\em Soft Matter\/} {\bf 17} 2120--2131

\bibitem{das2021coil}
Das S, Kennedy N and Cacciuto A 2021 {\em Soft Matter\/} {\bf 17} 160--164

\bibitem{anand2019beating}
Anand S~K, Chelakkot R and Singh S~P 2019 {\em Soft matter\/} {\bf 15} 7926--7933

\bibitem{clopes2021flagellar}
Clop{\'e}s J and Winkler R~G 2021 {\em The European Physical Journal E\/} {\bf 44} 1--12

\bibitem{bianco2018globulelike}
Bianco V, Locatelli E and Malgaretti P 2018 {\em Physical review letters\/} {\bf 121} 217802

\bibitem{fily2020buckling}
Fily Y, Subramanian P, Schneider T~M, Chelakkot R and Gopinath A 2020 {\em Journal of the Royal Society Interface\/} {\bf 17} 20190794

\bibitem{harada1987sliding}
Harada Y, Noguchi A, Kishino A and Yanagida T 1987 {\em Nature\/} {\bf 326} 805--808

\bibitem{heeremans2022chromatographic}
Heeremans T, Deblais A, Bonn D and Woutersen S 2022 {\em Science Advances\/} {\bf 8} eabj7918

\bibitem{manna2019emergent}
Manna R~K and Kumar P~S 2019 {\em Soft Matter\/} {\bf 15} 477--486

\bibitem{brennen1977fluid}
Brennen C and Winet H 1977 {\em Annual Review of Fluid Mechanics\/} {\bf 9} 339--398

\bibitem{howard2002mechanics}
Howard J and Clark R 2002 {\em Appl. Mech. Rev.\/} {\bf 55} B39--B39

\bibitem{ndlec1997self}
Ndlec F, Surrey T, Maggs A~C and Leibler S 1997 {\em Nature\/} {\bf 389} 305--308

\bibitem{nishiguchi2018flagellar}
Nishiguchi D, Iwasawa J, Jiang H~R and Sano M 2018 {\em New Journal of Physics\/} {\bf 20} 015002

\bibitem{deblais2020phase}
Deblais A, Maggs A, Bonn D and Woutersen S 2020 {\em Physical Review Letters\/} {\bf 124} 208006

\bibitem{anand2019behavior}
Anand S~K and Singh S~P 2019 {\em Soft Matter\/} {\bf 15} 4008--4018

\bibitem{biswas2017linking}
Biswas B, Manna R~K, Laskar A, Kumar P~S, Adhikari R and Kumaraswamy G 2017 {\em ACS nano\/} {\bf 11} 10025--10031

\bibitem{anand2018structure}
Anand S~K and Singh S~P 2018 {\em Physical Review E\/} {\bf 98} 042501

\bibitem{anand2020conformation}
Anand S~K and Singh S~P 2020 {\em Physical Review E\/} {\bf 101} 030501

\bibitem{vatin2024conformation}
Vatin M, Kundu S and Locatelli E 2024 {\em Soft Matter\/} {\bf 20} 1892--1904

\bibitem{natali2020local}
Natali L, Caprini L and Cecconi F 2020 {\em Soft Matter\/} {\bf 16} 2594--2604

\bibitem{vandebroek2015dynamics}
Vandebroek H and Vanderzande C 2015 {\em Physical Review E\/} {\bf 92} 060601

\bibitem{ghosh2014dynamics}
Ghosh A and Gov N 2014 {\em Biophysical journal\/} {\bf 107} 1065--1073

\bibitem{prathyusha2018dynamically}
Prathyusha K, Henkes S and Sknepnek R 2018 {\em Physical Review E\/} {\bf 97} 022606

\bibitem{laskar2017filament}
Laskar A and Adhikari R 2017 {\em New Journal of Physics\/} {\bf 19} 033021

\bibitem{chelakkot2014flagellar}
Chelakkot R, Gopinath A, Mahadevan L and Hagan M~F 2014 {\em Journal of The Royal Society Interface\/} {\bf 11} 20130884

\bibitem{ravichandran2017enhanced}
Ravichandran A, Vliegenthart G~A, Saggiorato G, Auth T and Gompper G 2017 {\em Biophysical journal\/} {\bf 113} 1121--1132

\bibitem{abkenar2013collective}
Abkenar M, Marx K, Auth T and Gompper G 2013 {\em Physical Review E\/} {\bf 88} 062314

\bibitem{foglino2019non}
Foglino M, Locatelli E, Brackley C, Michieletto D, Likos C and Marenduzzo D 2019 {\em Soft matter\/} {\bf 15} 5995--6005

\bibitem{eisenstecken2016conformational}
Eisenstecken T, Gompper G and Winkler R~G 2016 {\em Polymers\/} {\bf 8} 304

\bibitem{eisenstecken2017internal}
Eisenstecken T, Gompper G and Winkler R~G 2017 {\em The Journal of chemical physics\/} {\bf 146}

\bibitem{samanta2016chain}
Samanta N and Chakrabarti R 2016 {\em Journal of Physics A: Mathematical and Theoretical\/} {\bf 49} 195601

\bibitem{paul2022activity}
Paul S, Majumder S and Janke W 2022 {\em Soft Matter\/} {\bf 18} 6392--6403

\bibitem{das2019deviations}
Das S and Cacciuto A 2019 {\em Physical review letters\/} {\bf 123} 087802

\bibitem{philipps2022tangentially}
Philipps C~A, Gompper G and Winkler R~G 2022 {\em The Journal of Chemical Physics\/} {\bf 157}

\bibitem{tejedor2024progressive}
Tejedor A~R, Ram{\'\i}rez J and Ripoll M 2024 {\em Physical Review Research\/} {\bf 6} L032002

\bibitem{fazelzadeh2023effects}
Fazelzadeh M, Irani E, Mokhtari Z and Jabbari-Farouji S 2023 {\em Physical Review E\/} {\bf 108} 024606

\bibitem{malgaretti2024coil}
Malgaretti P, Locatelli E and Valeriani C 2024 {\em Molecular Physics\/}  e2384462

\bibitem{lamura2024excluded}
Lamura A 2024 {\em Physical Review E\/} {\bf 109} 054611

\bibitem{isele2015self}
Isele-Holder R~E, Elgeti J and Gompper G 2015 {\em Soft matter\/} {\bf 11} 7181--7190

\bibitem{gupta2019morphological}
Gupta N, Chaudhuri A and Chaudhuri D 2019 {\em Physical Review E\/} {\bf 99} 042405

\bibitem{D4SM00511B}
Karan C, Chaudhuri A and Chaudhuri D 2024 {\em Soft Matter\/} {\bf 20}(31) 6221--6230

\bibitem{kaiser2015does}
Kaiser A, Babel S, ten Hagen B, von Ferber C and L{\"o}wen H 2015 {\em The Journal of chemical physics\/} {\bf 142}

\bibitem{khalilian2024structural}
Khalilian H, Peruani F and Sarabadani J 2024 {\em arXiv preprint arXiv:2401.01719\/}

\bibitem{chaki2019enhanced}
Chaki S and Chakrabarti R 2019 {\em The Journal of chemical physics\/} {\bf 150}

\bibitem{goswami2022reconfiguration}
Goswami K, Chaki S and Chakrabarti R 2022 {\em Journal of Physics A: Mathematical and Theoretical\/} {\bf 55} 423002

\bibitem{oliver2018synchronization}
Oliver N, Alpmann C, Barroso {\'A}, Dewenter L, Woerdemann M and Denz C 2018 {\em Soft Matter\/} {\bf 14} 3073--3077

\bibitem{zhang2020reconfigurable}
Zhang B, Sokolov A and Snezhko A 2020 {\em Nature communications\/} {\bf 11} 4401

\bibitem{kreienkamp2022clustering}
Kreienkamp K~L and Klapp S~H 2022 {\em New Journal of Physics\/} {\bf 24} 123009

\bibitem{liu2019configuration}
Liu X, Jiang H and Hou Z 2019 {\em The Journal of chemical physics\/} {\bf 151}

\bibitem{li2023chirality}
Li J~R, Zhu W~j, Li J~J, Wu J~C and Ai B~Q 2023 {\em New Journal of Physics\/} {\bf 25} 043031

\bibitem{liebchen2022chiral}
Liebchen B and Levis D 2022 {\em Europhysics Letters\/} {\bf 139} 67001

\bibitem{lauga2006swimming}
Lauga E, DiLuzio W~R, Whitesides G~M and Stone H~A 2006 {\em Biophysical journal\/} {\bf 90} 400--412

\bibitem{kraft2013brownian}
Kraft D~J, Wittkowski R, Ten~Hagen B, Edmond K~V, Pine D~J and L{\"o}wen H 2013 {\em Physical Review E—Statistical, Nonlinear, and Soft Matter Physics\/} {\bf 88} 050301

\bibitem{aubret2018targeted}
Aubret A, Youssef M, Sacanna S and Palacci J 2018 {\em Nature Physics\/} {\bf 14} 1114--1118

\bibitem{vutukuri2017rational}
Vutukuri H~R, Bet B, Van~Roij R, Dijkstra M and Huck W~T 2017 {\em Scientific reports\/} {\bf 7} 16758

\bibitem{wykes2016dynamic}
Wykes M~S~D, Palacci J, Adachi T, Ristroph L, Zhong X, Ward M~D, Zhang J and Shelley M~J 2016 {\em Soft matter\/} {\bf 12} 4584--4589

\bibitem{narinder2018memory}
Narinder N, Bechinger C and Gomez-Solano J~R 2018 {\em Physical review letters\/} {\bf 121} 078003

\bibitem{kummel2013circular}
K{\"u}mmel F, Ten~Hagen B, Wittkowski R, Buttinoni I, Eichhorn R, Volpe G, L{\"o}wen H and Bechinger C 2013 {\em Physical review letters\/} {\bf 110} 198302

\bibitem{caprini2019active}
Caprini L and Marconi U~M~B 2019 {\em Soft matter\/} {\bf 15} 2627--2637

\bibitem{arora2021emergent}
Arora P, Sood A and Ganapathy R 2021 {\em Science Advances\/} {\bf 7} eabd0331

\bibitem{caprini2023chiral}
Caprini L, L{\"o}wen H and Marconi U~M~B 2023 {\em Soft Matter\/} {\bf 19} 6234--6246

\bibitem{zhou2023dynamics}
Zhou X, Wang Y, Xu B, Liu Y, Lu D, Luo J and Yang Z 2023 {\em AIP Advances\/} {\bf 13}

\bibitem{anand2024computer}
Anand S~K 2024 {\em The Journal of Chemical Physics\/} {\bf 161}

\bibitem{lamura2022self}
Lamura A 2022 {\em Polymers\/} {\bf 14} 4762

\bibitem{anand2021migration}
Anand S~K and Singh S~P 2021 {\em The European Physical Journal E\/} {\bf 44} 1--10

\bibitem{kaiser2014unusual}
Kaiser A and L{\"o}wen H 2014 {\em The Journal of chemical physics\/} {\bf 141}

\bibitem{harder2014activity}
Harder J, Valeriani C and Cacciuto A 2014 {\em Physical Review E\/} {\bf 90} 062312

\bibitem{anderson2022polymer}
Anderson C~J, Briand G, Dauchot O and Fern{\'a}ndez-Nieves A 2022 {\em Physical Review E\/} {\bf 106} 064606

\bibitem{Note1}
See Supplemental Material at URL-will-be-inserted-by-publisher.

\bibitem{sabeur2008kinetically}
Sabeur S~A, Hamdache F and Schmid F 2008 {\em Physical Review E—Statistical, Nonlinear, and Soft Matter Physics\/} {\bf 77} 020802

\bibitem{mitra2023emergent}
Mitra D and Chatterji A 2023 {\em arXiv preprint arXiv:2304.05633\/}

\bibitem{caprini2024spontaneous}
Caprini L, Abdoli I, Marconi U~M~B and L{\"o}wen H 2024 {\em arXiv preprint arXiv:2410.02567\/}

\bibitem{martin2020hydrodynamics}
Martin-Gomez A, Eisenstecken T, Gompper G and Winkler R~G 2020 {\em Physical Review E\/} {\bf 101} 052612

\bibitem{martin2019active}
Mart{\'\i}n-G{\'o}mez A, Eisenstecken T, Gompper G and Winkler R~G 2019 {\em Soft matter\/} {\bf 15} 3957--3969

\bibitem{liu2018individual}
Liu L, Chen J and An L 2018 {\em The Journal of Chemical Physics\/} {\bf 149}

\bibitem{singh2020flow}
Singh S~P and Winkler R~G 2020 {\em Journal of Rheology\/} {\bf 64} 1121--1131

\end{thebibliography}
\end{document}